\documentclass[12pt,a4paper]{revtex4}
\usepackage[utf8x]{inputenc}
\usepackage{ucs}
\usepackage[english]{babel}
\usepackage{amsmath}
\usepackage{amsfonts}
\usepackage{amssymb}
\usepackage{graphicx}
\usepackage[left=2cm,right=2cm,top=2cm,bottom=2cm]{geometry}

\usepackage{subfig}

\begin{document}
{
\title{Energy spectrum and quantum phase transition of the coupled single spin and an infinitely coordinated Ising chain}

\author{S.S. Seidov}
\affiliation{HSE University, Moscow, Russia}
\affiliation{NUST MISIS, Moscow, Russia}
\author{N.G. Pugach}
\affiliation{HSE University, Moscow, Russia}
\author{A.S. Sidorenko}
\affiliation{Technical University of Moldova, Institute of Electronic Engineering and Nanotechnologies, Republic of Moldova}

\begin{abstract}
We consider a spin model, composed of a single spin, connected to an infinitely coordinated Ising chain. Theoretical models of this type arise in various fields of theoretical physics, such as theory of open systems, quantum control and quantum computations. In the thermodynamic limit of infinite chain, we map the chain Hamiltonian to the Hamiltonian of the Lipkin–Meshkov–Glik model and the system as a whole is described by a generalized Rabi Hamiltonian. Next the effective Hamiltonian is obtained using Foulton–Gouterman transformation. In thermodynamic limit we obtain the spectrum of the whole system and study the properties of the ground state quantum phase transition.
\end{abstract}
\maketitle


\section{Introduction}
In the present manuscript we consider a single spin, connected to an infinitely coordinated Ising chain. From purely theoretical point of view, this model arises when studying the physics of open systems \cite{prokofev_theory_2000, arenz_control_2014}. In this case the chain is modelling the external environment, to which the single spin is connected. In such models it is convenient to study not only Markovian dynamics of the single spin, but also the non--Markovian one going beyond the limitations of the Lindblad master equation \cite{krovi_non-markovian_2007, budini_conditional_2019, han_non-markovianity_2020, brenes_bath-induced_2024}. The approach is to find the dynamics of the whole system, i.e. the chain and the single spin, and then trace out the chain degrees of freedom, ending up with the master equation for the single spin density matrix. One might choose to make or not to make the Markov approximation, obtaining different types of the master equations. Given that the exact solution is known, different master equations solutions can be compared against it. This allows to study the limits of applicability of the Markovian approximation and also the correct way of introducing the Lindblad dissipation operators. Said problem remains important in the general field of open quantum systems, extending beyond the spin models \cite{cattaneo_local_2019, mitchison_non-additive_2018, czerwinski_dynamics_2022}. 

One of the practical applications is modelling of certain quantum computing layouts, if one considers spins as qubits. In particular, previously we have proposed a method of implementing a CCZ (control--control--Z) quantum gate on a system, composed of three logical qubits, which are connected to another coupler--qubit \cite{simakov_high-fidelity_2024}. This approach allows to increase the fidelity of the operation, as well as has technical benefits such as simplicity of calibration and suppression of the unwanted longitudinal ZZ interaction. One of the important quantities is the shift of the coupler qubits energy levels depending on the state of the logical qubits. In the present manuscript we find the energy levels of such system in the limit of infinitely many logical qubits and find the energy spectrum of the coupler qubit depending on the state of the logical qubits ensemble.

We start our theoretical analysis by mapping the Ising chain Hamiltonian to a Lipkin--Meshkov--Glik (LMG) Hamiltonian \cite{lipkin_validity_1965, ribeiro_exact_2008, chinni_effect_of_chaos}. The Hamiltonian of the whole system then becomes akin to the Hamiltonian of the generalized Rabi model, but with bosonic field replaced by the collective spin of the LMG model. Next it is diagonalized in the spin space using Fulton--Gouterman transformation and we obtain an effective Hamiltonian. In the limit of infinite Ising chain, or equivalently of the infinite total spin, the LMG Hamiltonian can be solved exactly. We exploit this fact and obtain analytically the energy spectrum of the whole system. As it is known, in thermodynamic limit the LMG Hamiltonian undergoes a phase transition between the symmetric and broken symmetry phases. Coupling to the external spin shifts the critical value of the parameters, at which the phase transition happens, as well as the properties of the ground state. Investigating the structure of the minima of the ground state we find corrections to the critical values of the phase transition due to coupling to the external spin. 

\section{Model}
We consider a single spin, coupled to a fully connected Ising chain, with the Hamiltonian
\begin{equation}
\begin{aligned}
&H = \frac{\omega}{2} \tau_z + \frac{\Delta}{2} \tau_x + H_\text{chain} + H_\text{int}\\
&H_\text{chain} = \frac{1}{2}\sum_{i=1}^N (\tilde \omega \sigma_z^i + \tilde \Delta \sigma_x^i) + \frac{J}{4N} \sum_{i \neq j}^N \sigma_z^i \sigma_z^j\\
&H_\text{int} = \frac{\tilde J}{2} \tau_z \sum_{i=1}^N \sigma_z^i.
\end{aligned}
\end{equation}
Here $\tau_{x,z}$ are the Pauli matrices, describing the single spin and $\sigma_{x,z}^i$ are the Pauli matrices, describing spins in the Ising chain. This model arises when one studies the spin--bath theoretical models, in studies of quantum control and design of qubit layouts in quantum computations. The coupling between the spins in the chain is rescaled by $1/N$ factor in order to obtain finite energy per spin $\left\langle H\right\rangle/N$ in the thermodynamic limit. 

Let us first consider the Hamiltonian $H_\text{chain} + H_\text{int}$. By introducing collective spin operators 
\begin{equation}
S_{x,z} = \frac{1}{2}\sum_{i = 1}^N \sigma_{x,z}^i,
\end{equation}
the Hamiltonian is brought in form \cite{santos_excited-state_2016, vidal_concurrence_2006}
\begin{equation}\label{eq:H_LMG_1}
H_\text{chain} + H_\text{int} = \left(\tilde{\omega}+\frac{\tilde{J}}{2}\tau_z \right)S_z+\tilde{\Delta}S_x+\frac{J}{2S}S_z^2.
\end{equation}
Here $S = N/2$ is the total spin of the chain. This is a well--known Lipkin--Meshkov--Glick (LMG) Hamiltonian which we will further denote as $H_\text{LMG} = H_\text{chain} + H_\text{int}$. The total Hamiltonian now can be written as a $2 \times 2$ block matrix in the single spin Hilbert space:
\begin{equation}\label{eq:H_block}
H = \frac{1}{2} \begin{pmatrix}
\omega & \Delta \\
\Delta &-\omega
\end{pmatrix} + 
\begin{pmatrix}
H_\text{LMG}^+ & 0\\
0 & H_\text{LMG}^-
\end{pmatrix}.
\end{equation}
Here $H_\text{LMG}^\pm$ are the Hamiltonians $H_\text{LMG}$ corresponding to eigenvalues $\pm 1$ of $\tau_z$. These types of Hamiltonians are the Hamiltonians of the generalized Rabi models: these describe a two--level system connected not to a single bosonic mode, but some more complicated environment \cite{moroz_generalized_2016, eckle_generalization_2017, li_generalized_2021}.

\section{Diagonalization in spin space}
The Hamiltonian in the spin space can be diagonalized using the formula for the determinant of a $2 \times 2$ block matrix. This is also known as the Fulton--Gouterman transformation \cite{naseri_unconventional_2020}. This leads to two effective Hamiltonians in the chain Hilbert space, corresponding to the state of the single spin. These are
\begin{equation}\label{eq:H_eff_mat}
\begin{aligned}
&H_\text{eff}^\pm = \pm \frac{\omega}{2} + H_\text{LMG}^\pm - \frac{\Delta^2}{4} G_\mp\\
&G_\pm = \left(\pm \frac{\omega}{2} + H_\text{LMG}^\pm - E \right)^{-1}.
\end{aligned}
\end{equation}
Operators $G_\pm$ are the Green functions of the Hamiltonians $\pm \omega/2 + H_\text{LMG}^\pm$. Both of these Hamiltonians contain full information about the system, so it is sufficient to consider only one of them. We will choose the Hamiltonian $H_\text{eff} = H_\text{eff}^+$ as the effective Hamiltonian.

Given the eigenenergies $\varepsilon_n^\pm$ and eigenstates $|n^\pm \rangle$ of the Hamiltonian $\pm \omega/2 + H_\text{LMG}^\pm$, the effective Hamiltonian can be written as
\begin{equation}\label{eq:H_eff}
H_\text{eff} = \sum_{n=1}^N \varepsilon_n^+|n^+\rangle \langle n^+| - \frac{\Delta^2}{4} \sum_{n=1}^N \frac{|n^-\rangle \langle n^-|}{\varepsilon_n^- - E}.
\end{equation}
The eigenenergies of the whole system are solutions of the equation $\lambda(E) = E$, where $\lambda(E)$ are the eigenvalues of $H_\text{eff}$. In principle, solutions of this equation are exactly the energy levels of the corresponding physical system. However, given that in practice analytical solution is impossible in most cases, a usual approach is to substitute some value of energy $E_0$ in the left hand side and look for corrections. Our approach will be to find some kind of relation between the Hamiltonians $H^+_\text{LMG}$ and $H^-_\text{LMG}$, which will allow us to express the eigenstates of one Hamiltonian via the eigenstates of the other. Then the equation $\lambda(E) = E$ will be quadratic with two solutions, corresponding to two states of the single spin.

\section{Limit of strong single spin--chain coupling}
We focus on the limit of large coupling between the single spin and the chain, i.e. large $\tilde J$. In practice it is realized if one couples the single spin to an ensemble of noninteracting spins and the interaction between spins in the ensemble is indirect via the external spin. In this case spins in the chain are mostly aligned along the $z$--axis due to the large $\tilde J \tau_z S_z$ term. Effectively, interaction with the single spin creates a strong magnetic field, parallel to the single spin direction. The perpendicular component of the ``magnetic field'' $\tilde \Delta S_x$ thus can be considered as small perturbation.

Formally this means, that we can divide the LMG Hamiltonian into the main part
\begin{equation}
H^0_\text{LMG} = \left(\tilde{\omega}+\frac{\tilde{J}}{2}\tau_z\right)S_z+\frac{J}{2S}S_z^2
\end{equation}  
and perturbation $V = \tilde \Delta S_x$. With standard perturbation theory approach we find the energy levels of $H_\text{LMG}$ up to second order in $\tilde \Delta$:
\begin{equation}
\begin{aligned}
&\mathcal{E}^{\pm}_\sigma = \mathcal{E}^{(0)}_{\pm, \sigma} + \sum_{\sigma' \neq \sigma} c_{\sigma \sigma'}\\
&\mathcal{E}^{(0)}_{\pm, \sigma} =  \left(\tilde{\omega}\pm\frac{\tilde{J}}{2}\right)\sigma+\frac{J}{2S}\sigma^2\\
&c_{\sigma \sigma'}^\pm = \tilde \Delta^2 \frac{|\langle \sigma'| S_x| \sigma \rangle|^2}{\mathcal{E}^{(0)}_{\pm, \sigma} - \mathcal{E}^{(0)}_{\pm, \sigma'}}.
\end{aligned}
\end{equation}
Here $S_z |\sigma \rangle = \sigma |\sigma \rangle$. Accordingly, the eigenstates are
\begin{equation}
|\psi_\sigma^\pm \rangle \approx |\sigma \rangle  + \sum_{\sigma' \neq \sigma} c_{\sigma \sigma'}^\pm|\sigma' \rangle.
\end{equation}
As discussed earlier, we aim to relate $H_\text{LMG}^+$ and $H_\text{LMG}^-$. Let us express the projectors on states $|\psi_\sigma^+ \rangle$ via projectors on $|\psi_\sigma^- \rangle$. Up to second order in $\tilde \Delta$
\begin{equation}
|\psi_\sigma^+ \rangle \langle \psi_\sigma^+| = |\psi_\sigma^- \rangle \langle \psi_\sigma^-| + \sum_{\sigma'} (c^+_{\sigma \sigma'} - c^-_{\sigma \sigma'})(|\psi_\sigma^- \rangle \langle \sigma| + |\sigma \rangle \langle \psi_\sigma^-|).
\end{equation}
The Hamiltonians $H_\text{LMG}^\pm$ now can be written as
\begin{equation}
\begin{aligned}
&H^+_\text{LMG} = \sum_\sigma \mathcal{E}_\sigma^+ |\psi_\sigma^- \rangle \langle \psi_\sigma^-| + \sum_{\sigma \sigma'} \mathcal{E}_\sigma^+ (c^+_{\sigma \sigma'} - c^-_{\sigma \sigma'})(|\psi_\sigma^- \rangle \langle \sigma| + |\sigma \rangle \langle \psi_\sigma^-|)\\
&H^-_\text{LMG} = \sum_\sigma \mathcal{E}_\sigma^- |\psi_\sigma^- \rangle \langle \psi_\sigma^-|.
\end{aligned}
\end{equation}
One can see, that the leading order $H_\text{LMG}^+$ is expressed via projectors on the eigenstates of $H_\text{LMG}^-$. The extra terms, when substituted in the effective Hamiltonian, will lead to higher order corrections and will be insignificant. Indeed, substituting in (\ref{eq:H_eff}) we find
\begin{equation}\label{eq:Heff_large_J1}
\begin{aligned}
H_\text{eff} &= \sum_\sigma \left(\frac{\omega}{2} + \mathcal{E}_\sigma^+ + \frac{\Delta^2}{4(\mathcal{E}_\sigma^- - \omega/2 - E)} \right) |\psi_\sigma^- \rangle \langle \psi_\sigma^-| +\\
&+ \sum_{\sigma \sigma'} \mathcal{E}_\sigma^+ (c^+_{\sigma \sigma'} - c^-_{\sigma \sigma'}) (|\psi_\sigma^- \rangle \langle \sigma| + |\sigma \rangle \langle \psi_\sigma^-|).
\end{aligned}
\end{equation} 
The first term is diagonal in basis $|\psi_\sigma^- \rangle$, so its contribution to the eigenvalues of the effective Hamiltonian eigenvalues will be second order in $\tilde \Delta$ (as it is the order to which we have expanded $\mathcal{E}_\sigma^\pm)$. The second term is second order in $\tilde \Delta$ and off--diagonal, so its contribution will be fourth order in $\tilde \Delta$. Thus, up to second order in $\tilde \Delta$ the energy $E$ of the whole system is defined by equation
\begin{equation}
\frac{\omega}{2} + \mathcal{E}_\sigma^+ + \frac{\Delta^2}{4(\mathcal{E}_\sigma^- - \omega/2 - E)} = E,
\end{equation}
from which follows
\begin{equation}\label{eq:E_large_J1}
E_\sigma^\pm = \frac{1}{2} \left(\mathcal{E}_\sigma^+ + \mathcal{E}_\sigma^- \pm \sqrt{(\omega + \mathcal{E}_\sigma^+ - \mathcal{E}_\sigma^-)^2 + \Delta^2} \right).
\end{equation}
Also from these calculations follows, that the eigenstates are $|\psi^-_\sigma\rangle$. One might wonder why there is no contribution from  $|\psi^+_\sigma\rangle$, given that our choice between expanding the Hamiltonian (\ref{eq:Heff_large_J1}) in $|\psi^-_\sigma\rangle \langle \psi^-_\sigma|$ or $|\psi^+_\sigma\rangle \langle \psi^+_\sigma|$ was arbitrary. In fact, there is indeed no difference between choosing one over the other, because $\langle \psi^-_\sigma| \psi^+_{\sigma'} \rangle = \delta_{\sigma \sigma'} + \mathcal{O}(\tilde \Delta^4)$.

We also note, that the same spectrum corresponds to the single spin Hamiltonian
\begin{equation}
h = \frac{1}{2}\begin{pmatrix}
\omega & \Delta\\
\Delta & -\omega
\end{pmatrix} + 
\begin{pmatrix}
\mathcal{E}_\sigma^+ & 0\\
0 & \mathcal{E}_\sigma^-
\end{pmatrix}.
\end{equation}
This Hamiltonian can be obtained if one replaces $H_\text{LMG}^\pm$ by their eigenvalues $\mathcal{E}_\sigma^\pm$ in (\ref{eq:H_block}). This is a Born--Oppenheimer approximation, in which the chain is considered to be a fast subsystem relative to the single spin. In particular, the energy of the spin chain is a contribution to the potential energy of the single spin.

\section{Phase transition in thermodynamic limit}
\subsection{Phase transition of the bare LMG model}
In thermodynamic classical limit the spin operators in the LMG model can be replaced by classical expectation values, i.e. $S_z = S \cos \theta$, $S_x = S \sin \theta \cos \varphi$, $S_y = S \sin \theta \sin \varphi$. The Hamiltonian then is replaced by its classical energy profile, which is defined according to ref. \cite{ribeiro_exact_2008} as
\begin{equation}\label{eq:LMG_en}
\varepsilon^\pm\left(\theta,\varphi\right)= \lim_{S \rightarrow \infty}{\frac{\left\langle H_{\mathrm{LMG}}^\pm\right\rangle}{S}}\ \ =\left(\tilde{\omega}\pm\frac{\tilde{J}}{2}\right)\cos{\theta}+\frac{J}{2}\cos^2\theta+\tilde{\Delta}\sin{\theta}\cos{\varphi}.
\end{equation}
The average is taken over a spin coherent state $\left|\theta,\varphi\right\rangle$. It is known, that the LMG Hamiltonian has two distinct phases in thermodynamic limit 
\cite{ribeiro_exact_2008, botet_size_1982, botet_large-size_1983, castanos_classical_2006}. The symmetric phase, in which $|\langle S_z \rangle| = S$, is realized when the linear in $S_z$ term in the Hamiltonian dominates over the quadratic one. In our particular case this means competition between the values of coefficients $\tilde \omega + \tilde J \tau_z/2$ and $J$ in the Hamiltonian (\ref{eq:H_LMG_1}). The second broken symmetry phase, in which the energy profile has two minima at $\langle S_z \rangle = \pm S_z^0$, is realized in the opposite case, when the $\sim S_z^2$ term dominates over the $\sim S_z$ term. These minima are degenerate if $\tilde \Delta = 0$, otherwise one is lower than another. The plot of the LMG model energy as function of the angle $\theta$ is presented in fig. \ref{fig:LMG_en}.

We wish to study the phase transition of the bare LMG model, i.e. decoupled from the external spin, and in the next section we will compare the results with ones for the LMG model coupled to the external spin. First we have to find the extrema of the LMG model energy $\varepsilon = \varepsilon^+(\tilde J = 0)$. They are defined by equations
\begin{equation}
\begin{aligned}
&\frac{\partial \varepsilon}{\partial \theta} = 0 \Rightarrow S \sin \theta (J S \cos\theta + \tilde \omega) - \tilde \Delta S \cos\theta \cos\varphi = 0\\
&\frac{\partial \varepsilon}{\partial \varphi} = 0 \Rightarrow S \tilde \Delta \sin\theta \sin \varphi = 0.
\end{aligned}
\end{equation}
One of the solutions is $\sin\theta = 0$ and $\cos\varphi = 0$, it corresponds to the symmetric phase in which $|\langle S_z \rangle| = |S\cos\theta| = S$. The second solution corresponds to $\sin \varphi = 0$ and 
\begin{equation}\label{eq:cos_bare}
\cos \theta + \frac{\tilde \omega \sin\theta}{J \sin\theta - \tilde \Delta} = 0.
\end{equation}
It describes the broken symmetry phase in which $\left|\left\langle S_z\right\rangle\right|\neq S$.

We find the critical values of the parameters, at which the phase transition occurs, by checking the stability of the symmetric phase. Namely, if it is stable, the $\sin \theta=0$ and $\cos \varphi=0$ extremum is a minimum in $\theta$ direction and the second derivative of $\varepsilon^+\left(\theta,\varphi\right)$ with respect to $\theta$ is positive. Otherwise said extremum is a maximum and the stable phase is the broken symmetry one. When carrying out the calculations we should choose the $\theta=\pi$ solution of the equation $\sin{\theta}=0$, because the $\sim S_z$ contribution to the energy is positive and the ground state corresponds to $S_z=Scos{\pi}=-S$. The $\theta= 0$ solution corresponds to the maximum of the energy profile. So, we find
\begin{equation}
\frac{\partial^2}{\partial\theta^2}\varepsilon^+\big|_{\theta=\pi,\varphi=0}=\tilde{\omega}-J.
\end{equation}
Thus, the broken symmetry phase exists (i.e., the expression above is negative) for $\tilde{\omega}<{\widetilde{\omega}}_c^0=J$. The projection of the spin on the $z$-–axis in the broken symmetry phase is $S_z^0=S\cos{\theta_0}$, where $\theta_0$ is the solution of the equation (\ref{eq:cos_bare}). If $\tilde{\Delta}=0$, the solutions are $\cos{\theta_0}=\tilde{\omega}/J=\tilde{\omega}/{\tilde{\omega}}_c^0$. The plots of the energy profile $\varepsilon^+\left(\theta,\varphi\right)$ in symmetric and broken symmetry phases are presented in fig. \ref{fig:LMG_en}.
\begin{figure}[h!]
\center\includegraphics[width = 0.5\textwidth]{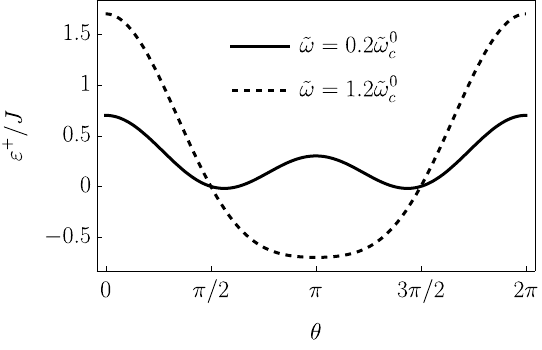}
\caption{Energy profile (\ref{eq:LMG_en}) of the LMG model in thermodynamic limit as function of $\theta$ and $\varphi = 0$. Solid line corresponds to the broken symmetry phase with two stable minima and the dashed line to the symmetric phase with a single minimum at  $\theta= \pi$.}
\label{fig:LMG_en}
\end{figure}

\subsection{Phase transition of the LMG model coupled to a single spin}
Now we study the properties of the phase transition if the chain is coupled to the external single spin. In this case we have to minimize the ground state energy of the whole system. From (\ref{eq:E_large_J1}) we find the spectrum
\begin{equation}\label{eq:E_total}
E^\pm(\theta, \varphi) = \frac{1}{2} \left(\varepsilon^+(\theta, \varphi) + \varepsilon^-(\theta, \varphi) \pm \sqrt{[\omega + \varepsilon^+(\theta, \varphi) - \varepsilon^-(\theta, \varphi)]^2 + \Delta^2} \right).
\end{equation}
These functions also have nontrivial minima structure, depending on the values of the parameters, see fig. \ref{fig:min_struct}. Again from equations $\partial_\theta E^- = 0$ and $\partial_\varphi E^- = 0$ we find that the extrema of the ground state energy are defined by equations
\begin{equation}\label{eq:FP_sym_brok}
\begin{aligned}
&\sin{\varphi}\sin{\theta}=0\\
&\left(\tilde{\Delta}\cos{\varphi}-J\sin{\theta}\right)\cos{\theta}-\tilde{\omega}\sin{\theta}=\frac{\tilde{J}\left(\omega+\tilde{J}\cos{\theta}\right)\sin{\theta}}{2\sqrt{\left(\omega+\tilde{J} \cos{\theta}\right)^2 + \Delta^2}}.
\end{aligned}
\end{equation}
The second equation defines $\left\langle S_z\right\rangle$ in the broken symmetry phase, analogously to equation (\ref{eq:cos_bare}). In general, it has up to nine real solutions on the interval $\theta\in\left[0,2\pi\right]$ depending on the values of the parameters. For $\tilde{\Delta}=0$ three of them are $\theta=0, \pi, 2\pi$ as it follows from the condition $\sin{\theta}=0$. Nonzero $\tilde{\Delta}$ will shift these solutions and corrections due to small $\tilde{\Delta}$ can be found by expanding the equation at said points. Six other solutions cannot be found analytically, however, we can study the properties of the energy profile exploiting the following facts:
\begin{enumerate}
\item Due to $2\pi$--periodicity of $E^\pm\left(\theta,\varphi\right)$ the extrema at $\theta=0, 2\pi$ are of the same type
\item Three of unknown extrema are on the interval $\theta\in\left(0,\pi\right)$ and the other three are on the interval $\theta\in\left(\pi,2\pi\right)$
\item A maximum should be followed by a minimum and vice versa
\end{enumerate}
Additional extrema arise due to hybridization between the energy levels $\varepsilon^\pm\left(\theta,\varphi\right)$ of the bare LMG-model with the single spin directed up or down, leading to appearance of avoided crossings and richer extremum structure of the ground state energy. The extrema at $\theta= 0, \pi, 2\pi$ always exist and could be either minima or maxima and the six other ones might be minima, maxima or not exist. This allows us to list all possible extrema configurations of the energy profile. We group them in to two types: either minima at $\theta= 0,2\pi$ and $\theta= \pi$ are of different type or the same, see fig. \ref{fig:min_struct}a and fig. \ref{fig:min_struct}b respectively.
\begin{figure}[h!]
\subfloat[]{\includegraphics[width=0.7\textwidth]{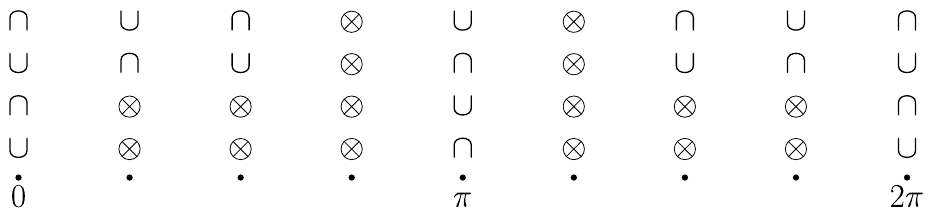}}\\
\subfloat[]{\includegraphics[width=0.7\textwidth]{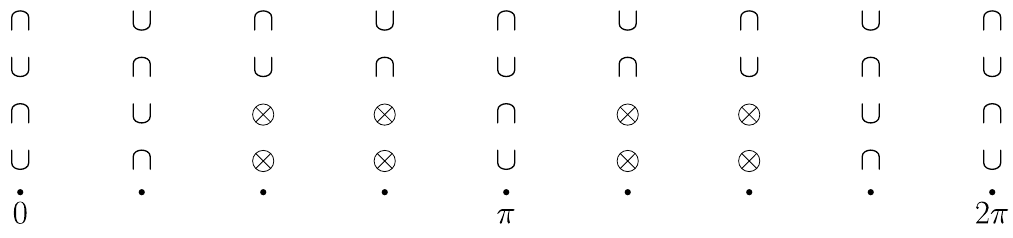}}
\caption{Possible configurations of the extrema of the ground state energy $E^-\left(\theta,\varphi\right)$. The symbol $\cup$ denotes a minimum, $\cap$ --- a maximum and $\otimes$ --- absence of the extremum. In (a) the ``different type'' configurations are presented and in (b) --- the ``same type''.}
\label{fig:min_struct}
\end{figure}

Let us first start with ``different type'' configurations in fig. \ref{fig:min_struct}a. Configurations 1 and 2 have no additional extrema and in both of them the symmetric phase is stable. The difference is, that in configuration 1 the total spin of the chain is aligned along the positive direction of the z-axis and in configuration 2 --- along the negative. Remarkably, configuration 1 is unstable for the bare LMG-model. Configuration 3 is not realized and in configuration 4 both symmetric and broken symmetry extrema are minima, meaning that one of the phases is stable, and the other is metastable. This means, that coupling to the external spin can change the type of the phase transition between two phases from the second to the first one. The corresponding plots of the energy levels are presented in fig. 3.
\begin{figure}[h!]
\subfloat[Configuration 1 from fig. \ref{fig:min_struct}a]{\includegraphics[width=0.49\textwidth]{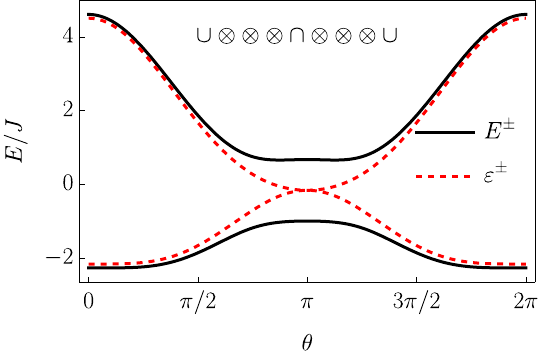}}
\subfloat[Configuration 2 from fig. \ref{fig:min_struct}a]{\includegraphics[width=0.49\textwidth]{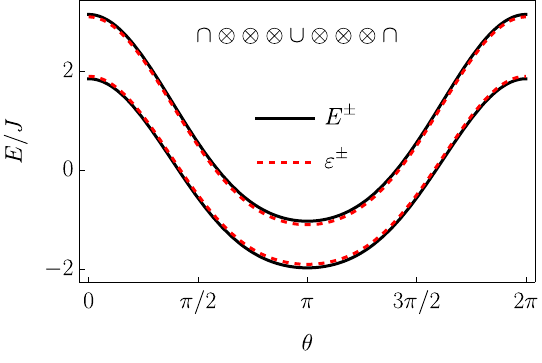}}\\
\subfloat[Configuration 4 from fig. \ref{fig:min_struct}a]{\includegraphics[width=0.49\textwidth]{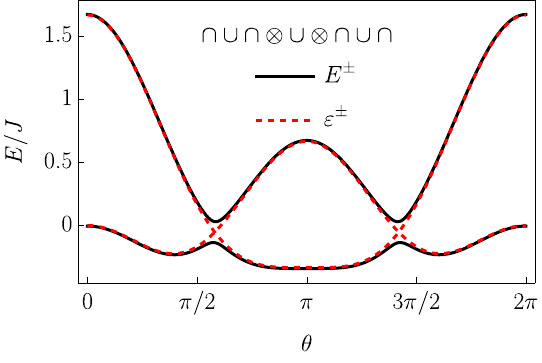}}
\caption{The energy levels of the system, corresponding to configurations 1, 2 and 4 from fig. \ref{fig:min_struct}a. In plots a) and b) the symmetric phases with spin aligned along positive and negative directions of the z-axis respectively are stable. In plot c) one can observe a stable symmetric phase and metastable broken symmetry phase. }
\label{fig:energies_a}
\end{figure}

Now we consider the ``same type'' configurations in fig. \ref{fig:min_struct}b. In configuration 1 two minima correspond to symmetric phases with total spin aligned along the positive and negative directions of the $z$--axis. One of the phases is stable and the other is metastable, so a first order phase transition between them is possible. In configuration 2 the broken symmetry phase is stable, resembling the case of the bare LMG--model. The configuration 3 is again not realized. The most interesting one is the configuration 4 in which the minima, corresponding to stable broken symmetry phase, split in to two. The plots of the energy levels, corresponding to described extrema configurations, can be found in fig. \ref{fig:energies_b}. 
\begin{figure}[h!]
\subfloat[Configuration 1 from fig. \ref{fig:min_struct}b]{\includegraphics[width=0.49\textwidth]{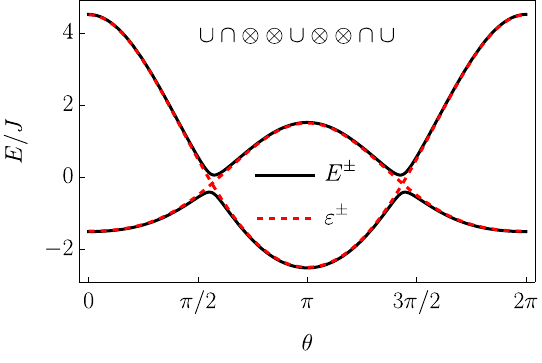}}
\subfloat[Configuration 2 from fig. \ref{fig:min_struct}b]{\includegraphics[width=0.49\textwidth]{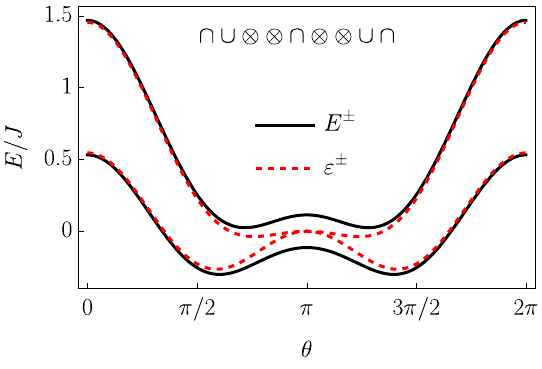}}\\
\subfloat[Configuration 4 from fig. \ref{fig:min_struct}b]{\includegraphics[width=0.49\textwidth]{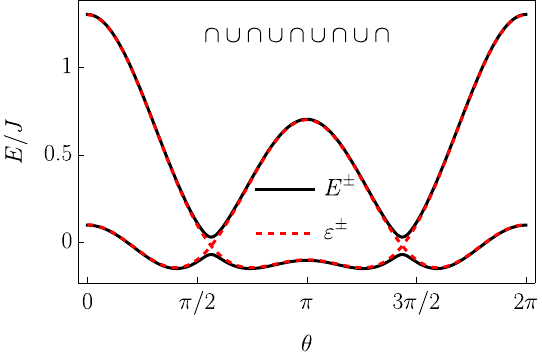}}
\caption{The energy levels of the system, corresponding to configurations 1, 2 and 4 from fig. \ref{fig:min_struct}b. In plot a) the symmetric phases with spin aligned along negative direction of the $z$--axis is stable and the one with spin aligned along the positive direction is metastable. In plot c) the minima, corresponding to stable broken symmetry phase are split into two.}
\label{fig:energies_b}
\end{figure}
We also derive the conditions for stability of the points  $\theta= 0, \pi, 2\pi$. Calculating the second derivatives we find
\begin{equation}
\begin{aligned}
&\frac{\partial^2}{\partial\theta^2}E^-\big|_{\theta=\pi}=\tilde{\omega}-\ J+\frac{\tilde{J}\left(\tilde{J}-\omega\right)}{2\sqrt{\left(\tilde{J}-\omega\right)^2+\Delta^2}}\\
&\frac{\partial^2}{\partial\theta^2}E^-\big|_{\theta=0,2\pi}=-\tilde{\omega}-\ J+\frac{\tilde{J}\left(\tilde{J}+\omega\right)}{2\sqrt{\left(\tilde{J}+\omega\right)^2+\Delta^2}}.
\end{aligned}
\end{equation}
Accordingly, the point $\theta=\pi$ is a minimum if
\begin{equation}\label{eq:pi_stable}
\tilde{\omega} > J-\frac{\widetilde{J}\left(\widetilde{J}-\omega\right)}{2\sqrt{\left(\widetilde{J}-\omega\right)^2+\Delta^2}}
\end{equation}
and points $\theta=0, 2\pi$ are stable if
\begin{equation}\label{eq:2pi_stable}
\tilde{\omega}<-J+\frac{\tilde{J}\left(\tilde{J}+\omega\right)}{2\sqrt{\left(\tilde{J}+\omega\right)^2+\Delta^2}}.
\end{equation}
Two important observations can be made here. First, the right-hand side of (\ref{eq:pi_stable}) can be negative for large enough $\tilde{J}$, i.e., single spin--chain coupling, making the symmetric state always stable. Second, at certain range of parameters both conditions (\ref{eq:pi_stable}) and (\ref{eq:2pi_stable}) might be true or not true simultaneously, which leads to appearance of the ``same type'' configurations. More detailed analysis of transitions between different phases requires knowing the conditions of existence intermediate extrema at $\theta\neq 0, \pi, 2\pi$. This boils down to finding all solution of the equation (\ref{eq:FP_sym_brok}) and one has to resort to numerical calculations. In fig. \ref{fig:diag} the phase diagram, obtained numerically, is presented on the $J$--$\tilde{J}$ plane for fixed values of the rest of the parameters.
\begin{figure}[h!]
\center\includegraphics[width = 0.5\textwidth]{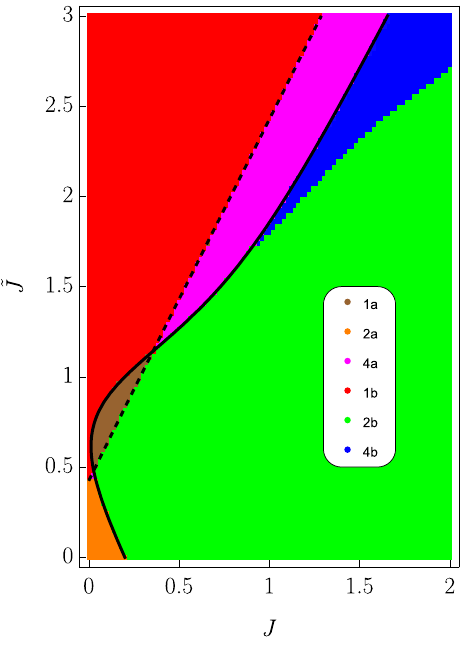}
\caption{The phase diagram of the considered model on the  $J$--$\tilde{J}$ plane. The legend notation coincides with the configurations in fig. \ref{fig:min_struct}, i.e., 1a is the configuration 1 in fig. \ref{fig:min_struct}a etc. Black solid line is the contour given by (\ref{eq:pi_stable}), the dashed black line is given by (\ref{eq:2pi_stable}). The rest of the parameters are chosen as $\omega = 1$, $\tilde{\omega}=0.2$, $\Delta=0.5$ and $\tilde{\Delta}=0$.}
\label{fig:diag}
\end{figure}

\section{Conclusions}
We have theoretically studied the infinitely coordinated Ising chain, coupled to a single external spin. We have written down the effective Hamiltonian in the Ising chain space by diagonalizing the Hamiltonian of the whole system in the space of the external spin. In thermodynamic limit, when the chains Hamiltonian is exactly solvable, the energy spectrum of the system was found. It is demonstrated, that coupling to an external spin drastically changes the properties of the phases relative to the bare LMG-model. In particular, it leads to appearance of a new stable phase --- symmetric one, but with the total spin of the chain oriented in the reverse direction compared to the symmetric state of the bare LMG--model. This phase might be stable as well as metastable and the first order phase transition between two stable phases is possible. The broken symmetry phase might also become metastable at certain values of the parameters, which means that the corresponding phase transition becomes of the first order as opposed to the second order in the case of the bare LMG--model. Finally, coupling to the external spin might change the properties of the broken symmetry phase by splitting the corresponding minima of energy in to two.

In case of experimental attempt, it is possible to check the spin configuration of the coordinated Ising chain, using a well-developed and very sensitive method of PNR (polarized neutron reflectometry) \cite{khaydukov_chirality_2021, khaydukov_proximity_2020, khaydukov_neutron_2019, zhaketov_structural_2023} to improve the qubit quality which is an important task \cite{klenov_examination_2008, klenov_superconductor-ferromagnet-superconductor_2010}.

The considered model is relevant in the field of quantum computations, as the layout is used for a certain realization of the CCZ gate \cite{simakov_high-fidelity_2024}. Although, admittedly, the thermodynamic limit approximation, made during the analysis is far from the experimentally feasible experimental setup involving only several qubits.

\section{Acknowledgements}
The publication was prepared with the support of the Russian Science Foundation (Grant No. 23--72--01067).


\begin{thebibliography}{28}
\expandafter\ifx\csname natexlab\endcsname\relax\def\natexlab#1{#1}\fi
\expandafter\ifx\csname bibnamefont\endcsname\relax
  \def\bibnamefont#1{#1}\fi
\expandafter\ifx\csname bibfnamefont\endcsname\relax
  \def\bibfnamefont#1{#1}\fi
\expandafter\ifx\csname citenamefont\endcsname\relax
  \def\citenamefont#1{#1}\fi
\expandafter\ifx\csname url\endcsname\relax
  \def\url#1{\texttt{#1}}\fi
\expandafter\ifx\csname urlprefix\endcsname\relax\def\urlprefix{URL }\fi
\providecommand{\bibinfo}[2]{#2}
\providecommand{\eprint}[2][]{\url{#2}}

\bibitem[{\citenamefont{Prokof'ev and Stamp}(2000)}]{prokofev_theory_2000}
\bibinfo{author}{\bibfnamefont{N.~V.} \bibnamefont{Prokof'ev}}
  \bibnamefont{and} \bibinfo{author}{\bibfnamefont{P.~C.~E.}
  \bibnamefont{Stamp}}, \bibinfo{journal}{Reports on Progress in Physics}
  \textbf{\bibinfo{volume}{63}}, \bibinfo{pages}{669} (\bibinfo{year}{2000}),
  ISSN \bibinfo{issn}{0034-4885, 1361-6633},
  \urlprefix\url{https://iopscience.iop.org/article/10.1088/0034-4885/63/4/204}.

\bibitem[{\citenamefont{Arenz et~al.}(2014)\citenamefont{Arenz, Gualdi, and
  Burgarth}}]{arenz_control_2014}
\bibinfo{author}{\bibfnamefont{C.}~\bibnamefont{Arenz}},
  \bibinfo{author}{\bibfnamefont{G.}~\bibnamefont{Gualdi}}, \bibnamefont{and}
  \bibinfo{author}{\bibfnamefont{D.}~\bibnamefont{Burgarth}},
  \bibinfo{journal}{New Journal of Physics} \textbf{\bibinfo{volume}{16}},
  \bibinfo{pages}{065023} (\bibinfo{year}{2014}), ISSN
  \bibinfo{issn}{1367-2630},
  \urlprefix\url{https://iopscience.iop.org/article/10.1088/1367-2630/16/6/065023}.

\bibitem[{\citenamefont{Krovi et~al.}(2007)\citenamefont{Krovi, Oreshkov,
  Ryazanov, and Lidar}}]{krovi_non-markovian_2007}
\bibinfo{author}{\bibfnamefont{H.}~\bibnamefont{Krovi}},
  \bibinfo{author}{\bibfnamefont{O.}~\bibnamefont{Oreshkov}},
  \bibinfo{author}{\bibfnamefont{M.}~\bibnamefont{Ryazanov}}, \bibnamefont{and}
  \bibinfo{author}{\bibfnamefont{D.~A.} \bibnamefont{Lidar}},
  \bibinfo{journal}{Physical Review A} \textbf{\bibinfo{volume}{76}},
  \bibinfo{pages}{052117} (\bibinfo{year}{2007}), ISSN
  \bibinfo{issn}{1050-2947, 1094-1622},
  \urlprefix\url{https://link.aps.org/doi/10.1103/PhysRevA.76.052117}.

\bibitem[{\citenamefont{Budini}(2019)}]{budini_conditional_2019}
\bibinfo{author}{\bibfnamefont{A.~A.} \bibnamefont{Budini}},
  \bibinfo{journal}{Physical Review A} \textbf{\bibinfo{volume}{99}},
  \bibinfo{pages}{052125} (\bibinfo{year}{2019}), ISSN
  \bibinfo{issn}{2469-9926, 2469-9934},
  \urlprefix\url{https://link.aps.org/doi/10.1103/PhysRevA.99.052125}.

\bibitem[{\citenamefont{Han et~al.}(2020)\citenamefont{Han, Zou, Li, and
  Shao}}]{han_non-markovianity_2020}
\bibinfo{author}{\bibfnamefont{L.}~\bibnamefont{Han}},
  \bibinfo{author}{\bibfnamefont{J.}~\bibnamefont{Zou}},
  \bibinfo{author}{\bibfnamefont{H.}~\bibnamefont{Li}}, \bibnamefont{and}
  \bibinfo{author}{\bibfnamefont{B.}~\bibnamefont{Shao}},
  \bibinfo{journal}{Entropy} \textbf{\bibinfo{volume}{22}},
  \bibinfo{pages}{895} (\bibinfo{year}{2020}), ISSN \bibinfo{issn}{1099-4300},
  \urlprefix\url{https://www.mdpi.com/1099-4300/22/8/895}.

\bibitem[{\citenamefont{Brenes et~al.}(2024)\citenamefont{Brenes, Min,
  Anto-Sztrikacs, Bar-Gill, and Segal}}]{brenes_bath-induced_2024}
\bibinfo{author}{\bibfnamefont{M.}~\bibnamefont{Brenes}},
  \bibinfo{author}{\bibfnamefont{B.}~\bibnamefont{Min}},
  \bibinfo{author}{\bibfnamefont{N.}~\bibnamefont{Anto-Sztrikacs}},
  \bibinfo{author}{\bibfnamefont{N.}~\bibnamefont{Bar-Gill}}, \bibnamefont{and}
  \bibinfo{author}{\bibfnamefont{D.}~\bibnamefont{Segal}},
  \bibinfo{journal}{The Journal of Chemical Physics}
  \textbf{\bibinfo{volume}{160}}, \bibinfo{pages}{244106}
  (\bibinfo{year}{2024}), ISSN \bibinfo{issn}{0021-9606, 1089-7690},
  \urlprefix\url{https://pubs.aip.org/jcp/article/160/24/244106/3299302/Bath-induced-interactions-and-transient-dynamics}.

\bibitem[{\citenamefont{Cattaneo et~al.}(2019)\citenamefont{Cattaneo, Giorgi,
  Maniscalco, and Zambrini}}]{cattaneo_local_2019}
\bibinfo{author}{\bibfnamefont{M.}~\bibnamefont{Cattaneo}},
  \bibinfo{author}{\bibfnamefont{G.~L.} \bibnamefont{Giorgi}},
  \bibinfo{author}{\bibfnamefont{S.}~\bibnamefont{Maniscalco}},
  \bibnamefont{and} \bibinfo{author}{\bibfnamefont{R.}~\bibnamefont{Zambrini}},
  \bibinfo{journal}{New Journal of Physics} \textbf{\bibinfo{volume}{21}},
  \bibinfo{pages}{113045} (\bibinfo{year}{2019}), ISSN
  \bibinfo{issn}{1367-2630},
  \urlprefix\url{https://iopscience.iop.org/article/10.1088/1367-2630/ab54ac}.

\bibitem[{\citenamefont{Mitchison and
  Plenio}(2018)}]{mitchison_non-additive_2018}
\bibinfo{author}{\bibfnamefont{M.~T.} \bibnamefont{Mitchison}}
  \bibnamefont{and} \bibinfo{author}{\bibfnamefont{M.~B.}
  \bibnamefont{Plenio}}, \bibinfo{journal}{New Journal of Physics}
  \textbf{\bibinfo{volume}{20}}, \bibinfo{pages}{033005}
  (\bibinfo{year}{2018}), ISSN \bibinfo{issn}{1367-2630},
  \urlprefix\url{https://iopscience.iop.org/article/10.1088/1367-2630/aa9f70}.

\bibitem[{\citenamefont{Czerwinski}(2022)}]{czerwinski_dynamics_2022}
\bibinfo{author}{\bibfnamefont{A.}~\bibnamefont{Czerwinski}},
  \bibinfo{journal}{Symmetry} \textbf{\bibinfo{volume}{14}},
  \bibinfo{pages}{1752} (\bibinfo{year}{2022}), ISSN \bibinfo{issn}{2073-8994},
  \urlprefix\url{https://www.mdpi.com/2073-8994/14/8/1752}.

\bibitem[{\citenamefont{Simakov et~al.}(2024)\citenamefont{Simakov, Mazhorin,
  Moskalenko, Seidov, and Besedin}}]{simakov_high-fidelity_2024}
\bibinfo{author}{\bibfnamefont{I.~A.} \bibnamefont{Simakov}},
  \bibinfo{author}{\bibfnamefont{G.~S.} \bibnamefont{Mazhorin}},
  \bibinfo{author}{\bibfnamefont{I.~N.} \bibnamefont{Moskalenko}},
  \bibinfo{author}{\bibfnamefont{S.~S.} \bibnamefont{Seidov}},
  \bibnamefont{and} \bibinfo{author}{\bibfnamefont{I.~S.}
  \bibnamefont{Besedin}}, \bibinfo{journal}{Physical Review Applied}
  \textbf{\bibinfo{volume}{21}}, \bibinfo{pages}{044035}
  (\bibinfo{year}{2024}), ISSN \bibinfo{issn}{2331-7019},
  \urlprefix\url{https://link.aps.org/doi/10.1103/PhysRevApplied.21.044035}.

\bibitem[{\citenamefont{Lipkin et~al.}(1965)\citenamefont{Lipkin, Meshkov, and
  Glick}}]{lipkin_validity_1965}
\bibinfo{author}{\bibfnamefont{H.~J.} \bibnamefont{Lipkin}},
  \bibinfo{author}{\bibfnamefont{N.}~\bibnamefont{Meshkov}}, \bibnamefont{and}
  \bibinfo{author}{\bibfnamefont{A.~J.} \bibnamefont{Glick}},
  \bibinfo{journal}{Nuclear Physics} \textbf{\bibinfo{volume}{62}},
  \bibinfo{pages}{188} (\bibinfo{year}{1965}), ISSN \bibinfo{issn}{0029-5582},
  \urlprefix\url{https://www.sciencedirect.com/science/article/pii/002955826590862X}.

\bibitem[{\citenamefont{Ribeiro et~al.}(2008)\citenamefont{Ribeiro, Vidal, and
  Mosseri}}]{ribeiro_exact_2008}
\bibinfo{author}{\bibfnamefont{P.}~\bibnamefont{Ribeiro}},
  \bibinfo{author}{\bibfnamefont{J.}~\bibnamefont{Vidal}}, \bibnamefont{and}
  \bibinfo{author}{\bibfnamefont{R.}~\bibnamefont{Mosseri}},
  \bibinfo{journal}{Physical Review E} \textbf{\bibinfo{volume}{78}},
  \bibinfo{pages}{021106} (\bibinfo{year}{2008}), ISSN
  \bibinfo{issn}{1539-3755, 1550-2376},
  \urlprefix\url{https://link.aps.org/doi/10.1103/PhysRevE.78.021106}.

\bibitem[{\citenamefont{Chinni et~al.}(2021)\citenamefont{Chinni, Poggi, and
  Deutsch}}]{chinni_effect_of_chaos}
\bibinfo{author}{\bibfnamefont{K.}~\bibnamefont{Chinni}},
  \bibinfo{author}{\bibfnamefont{P.~M.} \bibnamefont{Poggi}}, \bibnamefont{and}
  \bibinfo{author}{\bibfnamefont{I.~H.} \bibnamefont{Deutsch}},
  \bibinfo{journal}{Phys. Rev. Res.} \textbf{\bibinfo{volume}{3}},
  \bibinfo{pages}{033145} (\bibinfo{year}{2021}),
  \urlprefix\url{https://link.aps.org/doi/10.1103/PhysRevResearch.3.033145}.

\bibitem[{\citenamefont{Santos et~al.}(2016)\citenamefont{Santos, Tavora, and
  Perez-Bernal}}]{santos_excited-state_2016}
\bibinfo{author}{\bibfnamefont{L.~F.} \bibnamefont{Santos}},
  \bibinfo{author}{\bibfnamefont{M.}~\bibnamefont{Tavora}}, \bibnamefont{and}
  \bibinfo{author}{\bibfnamefont{F.}~\bibnamefont{Perez-Bernal}},
  \bibinfo{journal}{Physical Review A} \textbf{\bibinfo{volume}{94}},
  \bibinfo{pages}{012113} (\bibinfo{year}{2016}), ISSN
  \bibinfo{issn}{2469-9926, 2469-9934},
  \urlprefix\url{https://link.aps.org/doi/10.1103/PhysRevA.94.012113}.

\bibitem[{\citenamefont{Vidal}(2006)}]{vidal_concurrence_2006}
\bibinfo{author}{\bibfnamefont{J.}~\bibnamefont{Vidal}},
  \bibinfo{journal}{Physical Review A} \textbf{\bibinfo{volume}{73}},
  \bibinfo{pages}{062318} (\bibinfo{year}{2006}), ISSN
  \bibinfo{issn}{1050-2947, 1094-1622},
  \urlprefix\url{https://link.aps.org/doi/10.1103/PhysRevA.73.062318}.

\bibitem[{\citenamefont{Moroz}(2016)}]{moroz_generalized_2016}
\bibinfo{author}{\bibfnamefont{A.}~\bibnamefont{Moroz}}, \bibinfo{journal}{EPL
  (Europhysics Letters)} \textbf{\bibinfo{volume}{113}}, \bibinfo{pages}{50004}
  (\bibinfo{year}{2016}), ISSN \bibinfo{issn}{0295-5075, 1286-4854},
  \urlprefix\url{https://iopscience.iop.org/article/10.1209/0295-5075/113/50004}.

\bibitem[{\citenamefont{Eckle and
  Johannesson}(2017)}]{eckle_generalization_2017}
\bibinfo{author}{\bibfnamefont{H.-P.} \bibnamefont{Eckle}} \bibnamefont{and}
  \bibinfo{author}{\bibfnamefont{H.}~\bibnamefont{Johannesson}},
  \bibinfo{journal}{Journal of Physics A: Mathematical and Theoretical}
  \textbf{\bibinfo{volume}{50}}, \bibinfo{pages}{294004}
  (\bibinfo{year}{2017}), ISSN \bibinfo{issn}{1751-8113, 1751-8121},
  \urlprefix\url{https://iopscience.iop.org/article/10.1088/1751-8121/aa785a}.

\bibitem[{\citenamefont{Li and Batchelor}(2021)}]{li_generalized_2021}
\bibinfo{author}{\bibfnamefont{Z.-M.} \bibnamefont{Li}} \bibnamefont{and}
  \bibinfo{author}{\bibfnamefont{M.~T.} \bibnamefont{Batchelor}},
  \bibinfo{journal}{Physical Review A} \textbf{\bibinfo{volume}{104}},
  \bibinfo{pages}{033712} (\bibinfo{year}{2021}), ISSN
  \bibinfo{issn}{2469-9926, 2469-9934},
  \urlprefix\url{https://link.aps.org/doi/10.1103/PhysRevA.104.033712}.

\bibitem[{\citenamefont{Naseri et~al.}(2020)\citenamefont{Naseri, Hu, and
  Luo}}]{naseri_unconventional_2020}
\bibinfo{author}{\bibfnamefont{A.}~\bibnamefont{Naseri}},
  \bibinfo{author}{\bibfnamefont{Y.}~\bibnamefont{Hu}}, \bibnamefont{and}
  \bibinfo{author}{\bibfnamefont{W.}~\bibnamefont{Luo}},
  \emph{\bibinfo{title}{Unconventional supersymmetric quantum mechanics in spin
  systems}} (\bibinfo{year}{2020}),
  \urlprefix\url{https://arxiv.org/abs/2012.00197}.

\bibitem[{\citenamefont{Botet et~al.}(1982)\citenamefont{Botet, Jullien, and
  Pfeuty}}]{botet_size_1982}
\bibinfo{author}{\bibfnamefont{R.}~\bibnamefont{Botet}},
  \bibinfo{author}{\bibfnamefont{R.}~\bibnamefont{Jullien}}, \bibnamefont{and}
  \bibinfo{author}{\bibfnamefont{P.}~\bibnamefont{Pfeuty}},
  \bibinfo{journal}{Physical Review Letters} \textbf{\bibinfo{volume}{49}},
  \bibinfo{pages}{478} (\bibinfo{year}{1982}), ISSN \bibinfo{issn}{0031-9007},
  \urlprefix\url{https://link.aps.org/doi/10.1103/PhysRevLett.49.478}.

\bibitem[{\citenamefont{Botet and Jullien}(1983)}]{botet_large-size_1983}
\bibinfo{author}{\bibfnamefont{R.}~\bibnamefont{Botet}} \bibnamefont{and}
  \bibinfo{author}{\bibfnamefont{R.}~\bibnamefont{Jullien}},
  \bibinfo{journal}{Physical Review B} \textbf{\bibinfo{volume}{28}},
  \bibinfo{pages}{3955} (\bibinfo{year}{1983}), ISSN \bibinfo{issn}{0163-1829},
  \urlprefix\url{https://link.aps.org/doi/10.1103/PhysRevB.28.3955}.

\bibitem[{\citenamefont{Castanos et~al.}(2006)\citenamefont{Castanos,
  Lopez-Pena, Hirsch, and Lopez-Moreno}}]{castanos_classical_2006}
\bibinfo{author}{\bibfnamefont{O.}~\bibnamefont{Castanos}},
  \bibinfo{author}{\bibfnamefont{R.}~\bibnamefont{Lopez-Pena}},
  \bibinfo{author}{\bibfnamefont{J.~G.} \bibnamefont{Hirsch}},
  \bibnamefont{and}
  \bibinfo{author}{\bibfnamefont{E.}~\bibnamefont{Lopez-Moreno}},
  \bibinfo{journal}{Physical Review B} \textbf{\bibinfo{volume}{74}},
  \bibinfo{pages}{104118} (\bibinfo{year}{2006}), ISSN
  \bibinfo{issn}{1098-0121, 1550-235X},
  \urlprefix\url{https://link.aps.org/doi/10.1103/PhysRevB.74.104118}.

\bibitem[{\citenamefont{Khaydukov et~al.}(2021)\citenamefont{Khaydukov, Lenk,
  Zdravkov, Morari, Keller, Sidorenko, Tagirov, Tidecks, Horn, and
  Keimer}}]{khaydukov_chirality_2021}
\bibinfo{author}{\bibfnamefont{Y.~N.} \bibnamefont{Khaydukov}},
  \bibinfo{author}{\bibfnamefont{D.}~\bibnamefont{Lenk}},
  \bibinfo{author}{\bibfnamefont{V.}~\bibnamefont{Zdravkov}},
  \bibinfo{author}{\bibfnamefont{R.}~\bibnamefont{Morari}},
  \bibinfo{author}{\bibfnamefont{T.}~\bibnamefont{Keller}},
  \bibinfo{author}{\bibfnamefont{A.~S.} \bibnamefont{Sidorenko}},
  \bibinfo{author}{\bibfnamefont{L.~R.} \bibnamefont{Tagirov}},
  \bibinfo{author}{\bibfnamefont{R.}~\bibnamefont{Tidecks}},
  \bibinfo{author}{\bibfnamefont{S.}~\bibnamefont{Horn}}, \bibnamefont{and}
  \bibinfo{author}{\bibfnamefont{B.}~\bibnamefont{Keimer}},
  \bibinfo{journal}{Physical Review B} \textbf{\bibinfo{volume}{104}},
  \bibinfo{pages}{174445} (\bibinfo{year}{2021}), ISSN
  \bibinfo{issn}{2469-9950, 2469-9969},
  \urlprefix\url{https://link.aps.org/doi/10.1103/PhysRevB.104.174445}.

\bibitem[{\citenamefont{Khaydukov et~al.}(2020)\citenamefont{Khaydukov, Putter,
  Guasco, Morari, Kim, Keller, Sidorenko, and
  Keimer}}]{khaydukov_proximity_2020}
\bibinfo{author}{\bibfnamefont{Y.}~\bibnamefont{Khaydukov}},
  \bibinfo{author}{\bibfnamefont{S.}~\bibnamefont{Putter}},
  \bibinfo{author}{\bibfnamefont{L.}~\bibnamefont{Guasco}},
  \bibinfo{author}{\bibfnamefont{R.}~\bibnamefont{Morari}},
  \bibinfo{author}{\bibfnamefont{G.}~\bibnamefont{Kim}},
  \bibinfo{author}{\bibfnamefont{T.}~\bibnamefont{Keller}},
  \bibinfo{author}{\bibfnamefont{A.}~\bibnamefont{Sidorenko}},
  \bibnamefont{and} \bibinfo{author}{\bibfnamefont{B.}~\bibnamefont{Keimer}},
  \bibinfo{journal}{Beilstein Journal of Nanotechnology}
  \textbf{\bibinfo{volume}{11}}, \bibinfo{pages}{1254} (\bibinfo{year}{2020}),
  ISSN \bibinfo{issn}{2190-4286},
  \urlprefix\url{https://www.beilstein-journals.org/bjnano/articles/11/109}.

\bibitem[{\citenamefont{Khaydukov et~al.}(2019)\citenamefont{Khaydukov,
  Kravtsov, Morari, Lenk, Mustafa, Kim, Trapp, Zhaketov, Proglyado, Zrdavkov
  et~al.}}]{khaydukov_neutron_2019}
\bibinfo{author}{\bibfnamefont{Y.}~\bibnamefont{Khaydukov}},
  \bibinfo{author}{\bibfnamefont{E.}~\bibnamefont{Kravtsov}},
  \bibinfo{author}{\bibfnamefont{R.}~\bibnamefont{Morari}},
  \bibinfo{author}{\bibfnamefont{D.}~\bibnamefont{Lenk}},
  \bibinfo{author}{\bibfnamefont{L.}~\bibnamefont{Mustafa}},
  \bibinfo{author}{\bibfnamefont{G.}~\bibnamefont{Kim}},
  \bibinfo{author}{\bibfnamefont{M.}~\bibnamefont{Trapp}},
  \bibinfo{author}{\bibfnamefont{V.}~\bibnamefont{Zhaketov}},
  \bibinfo{author}{\bibfnamefont{V.}~\bibnamefont{Proglyado}},
  \bibinfo{author}{\bibfnamefont{V.}~\bibnamefont{Zrdavkov}},
  \bibnamefont{et~al.}, \bibinfo{journal}{Journal of Physics: Conference
  Series} \textbf{\bibinfo{volume}{1389}}, \bibinfo{pages}{012060}
  (\bibinfo{year}{2019}), ISSN \bibinfo{issn}{1742-6588, 1742-6596},
  \urlprefix\url{https://iopscience.iop.org/article/10.1088/1742-6596/1389/1/012060}.

\bibitem[{\citenamefont{Zhaketov et~al.}(2023)\citenamefont{Zhaketov,
  Devyaterikov, Avdeev, Norov, Kolupaev, Kuzmenko, Pugach, Khaydukov, Kravtsov,
  Nikitenko et~al.}}]{zhaketov_structural_2023}
\bibinfo{author}{\bibfnamefont{V.}~\bibnamefont{Zhaketov}},
  \bibinfo{author}{\bibfnamefont{D.}~\bibnamefont{Devyaterikov}},
  \bibinfo{author}{\bibfnamefont{M.}~\bibnamefont{Avdeev}},
  \bibinfo{author}{\bibfnamefont{D.}~\bibnamefont{Norov}},
  \bibinfo{author}{\bibfnamefont{E.}~\bibnamefont{Kolupaev}},
  \bibinfo{author}{\bibfnamefont{M.}~\bibnamefont{Kuzmenko}},
  \bibinfo{author}{\bibfnamefont{N.}~\bibnamefont{Pugach}},
  \bibinfo{author}{\bibfnamefont{Y.}~\bibnamefont{Khaydukov}},
  \bibinfo{author}{\bibfnamefont{E.}~\bibnamefont{Kravtsov}},
  \bibinfo{author}{\bibfnamefont{Y.}~\bibnamefont{Nikitenko}},
  \bibnamefont{et~al.}, \bibinfo{journal}{Physics of the Solid State}
  \textbf{\bibinfo{volume}{65}}, \bibinfo{pages}{1076} (\bibinfo{year}{2023}),
  ISSN \bibinfo{issn}{1063-7834, 1090-6460},
  \urlprefix\url{https://doi.org/10.61011/PSS.2023.07.56393.35H}.

\bibitem[{\citenamefont{Klenov et~al.}(2008)\citenamefont{Klenov, Kornev,
  Vedyayev, Ryzhanova, Pugach, and Rumyantseva}}]{klenov_examination_2008}
\bibinfo{author}{\bibfnamefont{N.}~\bibnamefont{Klenov}},
  \bibinfo{author}{\bibfnamefont{V.}~\bibnamefont{Kornev}},
  \bibinfo{author}{\bibfnamefont{A.}~\bibnamefont{Vedyayev}},
  \bibinfo{author}{\bibfnamefont{N.}~\bibnamefont{Ryzhanova}},
  \bibinfo{author}{\bibfnamefont{N.}~\bibnamefont{Pugach}}, \bibnamefont{and}
  \bibinfo{author}{\bibfnamefont{T.}~\bibnamefont{Rumyantseva}},
  \bibinfo{journal}{Journal of Physics: Conference Series}
  \textbf{\bibinfo{volume}{97}}, \bibinfo{pages}{012037}
  (\bibinfo{year}{2008}), ISSN \bibinfo{issn}{1742-6596},
  \urlprefix\url{https://iopscience.iop.org/article/10.1088/1742-6596/97/1/012037}.

\bibitem[{\citenamefont{Klenov et~al.}(2010)\citenamefont{Klenov, Kornev,
  Sharafiev, Bakurskiy, and
  Pugach}}]{klenov_superconductor-ferromagnet-superconductor_2010}
\bibinfo{author}{\bibfnamefont{N.~V.} \bibnamefont{Klenov}},
  \bibinfo{author}{\bibfnamefont{V.~K.} \bibnamefont{Kornev}},
  \bibinfo{author}{\bibfnamefont{A.~V.} \bibnamefont{Sharafiev}},
  \bibinfo{author}{\bibfnamefont{S.~V.} \bibnamefont{Bakurskiy}},
  \bibnamefont{and} \bibinfo{author}{\bibfnamefont{N.~G.}
  \bibnamefont{Pugach}}, \bibinfo{journal}{Journal of Physics: Conference
  Series} \textbf{\bibinfo{volume}{234}}, \bibinfo{pages}{042017}
  (\bibinfo{year}{2010}), ISSN \bibinfo{issn}{1742-6596},
  \urlprefix\url{https://iopscience.iop.org/article/10.1088/1742-6596/234/4/042017}.

\end{thebibliography}

\end{document}